**Title:** An analytical approach for calculating transfer integrals in superexchange coupled dimers

**Authors:** Stefan Lebernegg, Georg Amthauer and Michael Grodzicki

**Address:** Department of Materials Engineering and Physics, University of Salzburg, Hellbrunnerstrasse 34, 5020 Salzburg, Austria

**Running title:** Analytical calculation of transfer integrals

**Corresponding authors address:**

Phone:  ++43-662-8044-5463

Fax: ++43-662-8044-622

Mail: stefan.lebernegg@sbg.ac.at

**Abstract.** An analytical expression for the transfer integral $H_{AB}$ between the localized magnetic orbitals in superexchange-coupled dimers as a function of the type of atoms and geometry of the molecule has been derived by explicitly including orbital interactions. It is shown that $H_{AB}$ plays the key role for the magnetic coupling constant $J$ in understanding magneto-structural correlations. The reliability and capability of this approach is confirmed by comparison with numerical electronic structure calculations in the local spin-density approximation on singly and doubly bridged Cu(II)-dimers with fluorine ligands. All results can be calculated and understood within the analytical formalism representing, therefore, a powerful tool for understanding the magneto-structural correlations and also for constructing magnetic orbitals analytically.

## I. INTRODUCTION

The isotropic magnetic interaction between localized spins at centres $A$ and $B$ may be described by the Heisenberg-Dirac-van Vleck Hamiltonian

$$H = -2J \cdot \hat{S}^A \hat{S}^B \tag{1}$$

where the (isotropic) Heisenberg coupling constant $J$ describes the strength and mode of the magnetic coupling. With regard to this sign convention $J$ is positive for parallel or ferromagnetic coupling and negative for antiparallel or antiferromagnetic alignment of the two spins. Besides the quantitative determination of the size and sign of $J$ in particular systems, the central purpose of a general theory of superexchange is to identify and deduce the various geometric and electronic factors responsible for the mode of interaction between the spins in exchange coupled systems [1]. The various attempts of determining $J$ can roughly be described as follows:

(i) Experimentally, $J$ can be derived from spectroscopic data as EPR spectra [2] or from temperature dependent magnetic susceptibility measurements [3]. On this level nothing can be said conclusively about the nature and origin of the magnetic interactions determining sign and size of $J$.

(ii) Phenomenologically, a number of empirical rules and correlations have been inferred from the observation of certain regularities. Examples are the Goodenough-Kanamori rules [4,5] interpreted later by Andersons theory of superexchange [6] or the frequently proposed exponential dependence of $J$ on the distance between the interacting metal centers.

(iii) Theoretically, $J$ can be determined by numerical electronic structure calculations, especially by methods based on density functional theory (DFT) [7] in combination with the broken symmetry formalism [8]. Even though the dependence of $J$ on certain parameters can be investigated by systematic numerical model calculations, it is almost impossible to derive manageable analytical expressions for understanding magneto-structural correlations on that high level of sophistication.

Accordingly, none of these three approaches enables the derivation of explicit formulas for $J$ as a function of geometrical parameters, electronic properties of the metal centers, nature of the bridging and terminal ligands etc. Moreover, both the size of the experimentally investigated real systems and the complex nature of the magnetic interactions in those systems make it difficult to establish an explicit connection between the molecular (geometric and electronic) structure and magnetic properties by deriving an analytical expression for $J$, and a systematic analysis of the dependence of $J$ on these factors has not yet been reported to the best of our knowledge. It is, thus, necessary to select model systems that are simple enough to allow an approximately analytical treatment but, at the same time, are close enough to real systems where experimental data are available for comparison. Such paradigmatic model systems are binuclear Cu(II) complexes with halogens, oxygen or OH groups as bridging ligands since they contain a single magnetic orbital per metal center and exhibit antiferromagnetic as well as ferromagnetic coupling depending on the nature and arrangement of the bridging and the terminal ligands [1,9]. Moreover, such systems are of great practical importance due to their occurrence in biological systems [10] and in high-$T_c$ superconductors [11].

For a dimer with one magnetic orbital per metal centre, as realized in Cu(II)-dimers, the coupling constant for superexchange may be approximated by [6, 12]:

$$J = K_{AB} - \frac{2\left(H_{AB}\right)^2}{U_{AA} - U_{AB}} \qquad (2)$$

where the two-electron exchange integral $K_{AB}$ is always positive, hence leading to ferromagnetic coupling. The second term represents the antiferromagnetic contribution and arises from the (weak) delocalization of the magnetic orbitals towards the bridging ligands [6]. Extensions of this first approximation have been discussed [13,14] and thoroughly been reviewed [15] but for systems with a single nondegenerate magnetic orbital per metal eq. (2) is known to be sufficiently accurate. Another class of problems arises when using eqs. (1) and (2) in combination with numerical calculations within the frame of density functional theory (DFT). The first one concerns the proper definition of localized spins entering the Heisenberg Hamiltonian, eq. (1). The various possibilities for such a definition have recently been discussed [16,17] and reviewed [18]. However, the investigations of this work aiming at deriving qualitative magneto-structural correlations do not depend on such definitions. More important might be another problem since eq.(2) originates from a configuration interaction (CI) analysis [12] and DFT and CI calculations consider the electron correlation in different ways. However, based on a spin polarization perturbation orbital theory Seo [19] pointed out that the structure of eq.(2) is preserved for localized MO's derived from a DFT calculation.

Among the various integrals occurring in eq. (2) both the ferromagnetic term and the denominator, i.e. the effective on-site electronic repulsion $U = U_{AA} - U_{AB}$, of the second term are expected to vary only slightly with geometry [20]. By contrast, the transfer integral $H_{AB}$, that is expected to be roughly a function of the overlap integral between the magnetic orbitals of the parent monomers forming the molecule or solid [21,22], strongly depends on geometry representing thus the key parameter for deriving and understanding magneto-structural correlations. Although the transfer integral may be obtained from numerical electronic structure calculations, the analysis of the various contributions to $H_{AB}$, as well as the extraction of those interactions dominating the magneto-structural correlations are usually complicated if possible at all, i.e. the detailed structure of the magnetic coupling mechanism remains hidden. Accordingly, this work aims at deriving an expression for $H_{AB}$ and $J$ which has a sound physical basis in the sense that its general functional form can be justified by theoretical reasoning while the system-dependent parameters have a well defined physical meaning insofar as they can be determined, at least in principle, from sufficiently accurate numerical calculations. Although the derivation is carried out for homonuclear transition metal dimers with a single unpaired electron per metal centre, the final result will be applicable, as well, to heteronuclear complexes with more than one unpaired electron per metal site.

After some basic definitions of magnetic orbitals and transfer integrals in the second section, different approaches for their analytical calculation are derived in section three. Afterwards, the analytical approaches are applied on several Cu-F model systems. In the last section the correlation between the transfer integral and $J$ is investigated in detail and an analytical expression for $J$ will be used for estimating the magnetic behavior of model complexes. The analytical transfer integrals and coupling constants will be compared with the results of two different numerical DFT-codes.

## II. GENERAL BASIS

In the configuration interaction analysis from which eq.(2) is derived the transfer integral $H_{AB}$ is defined as [23]

$$H_{AB} = \langle \psi_A | H | \psi_B \rangle = \langle \psi_A(1) | h(1) | \psi_B(1) \rangle + \langle \psi_A(1)\psi_B(2) \| r_{12} \|^{-1} | \psi_B(1)\psi_B(2) \rangle \qquad (3)$$

where $H$ is the Fock-operator. $h(1)$ is the one-electron Hamiltonian and $\psi_A$ and $\psi_B$ are singly occupied molecular orbitals (MO's) of the dimer localized on the different metal centres $A$ and $B$. These orbitals contain the active electrons. The two-electron (bondcharge) integral is generally regarded as negligible [12, 8,24]. As already mentioned Seo [19] pointed out that the structure of eq.(2) is preserved for localized MO's derived from a DFT calculation where the transfer integral is defined as the expectation value of a spin-restricted DFT-Hamiltonian with energy-localized molecular orbitals containing the active electrons. These localized MO's are constructed via linearly combining the delocalized highest occupied (HOMO) and the lowest unoccupied (LUMO) Kohn-Sham orbitals of the dimer, denoted as $\psi_+$ and $\psi_-$,

$$|\psi_A\rangle = \cos\gamma \cdot |\psi_-\rangle + \sin\gamma \cdot |\psi_+\rangle \qquad |\psi_B\rangle = -\sin\gamma \cdot |\psi_-\rangle + \cos\gamma \cdot |\psi_+\rangle \qquad (4)$$

where $\gamma$ has to be estimated numerically in order to maximally localize the orbitals. For symmetric dimers eq.(4) reduces to

$$|\psi_A\rangle = 2^{-1/2}|\psi_+ + \psi_-\rangle \qquad |\psi_B\rangle = 2^{-1/2}|\psi_+ - \psi_-\rangle \qquad (5)$$

Accordingly, the transfer integral is given as

$$H_{AB} = \langle \psi_A | H | \psi_B \rangle = \frac{1}{2}\langle \psi_+ + \psi_- | H | \psi_+ - \psi_- \rangle = \frac{1}{2}(\varepsilon_+ - \varepsilon_-) \qquad (6)$$

where $\varepsilon_+$ and $\varepsilon_-$ are the orbital energies of $\psi_+$ and $\psi_-$. For symmetrical, planar, doubly bridged transition metal (TM)-dimer these orbitals are depicted in figure 1 B,C.

**FIG 1.**

The aim is to construct localized dimer MO's, $\left|\psi_A\right\rangle$ and $\left|\psi_B\right\rangle$, analytically in order to derive an analytical expression for the transfer integral and correspondingly for the coupling constant $J$. To this end, Anderson [6] suggested, in a first step, to solve the parent monomer problems, i.e. to construct the singly occupied MO of each of the two monomers (figure 1A), which, in a second step, are linearly combined to give the HOMO and LUMO of the dimer.

$$\left|\psi_\pm\right\rangle = N_\pm \left(\left|\psi_A^{mono}\right\rangle \pm \left|\psi_B^{mono}\right\rangle\right) = 2^{-0.5}\left(1 \pm S_{AB}^{mono}\right)^{-0.5}\left(\left|\psi_A^{mono}\right\rangle \pm \left|\psi_B^{mono}\right\rangle\right) \qquad (7)$$

with the overlap matrix element $S_{AB}^{mono} = \left\langle\psi_A^{mono}\middle|\psi_B^{mono}\right\rangle$. Combining the dimer-orbitals $\left|\psi_+\right\rangle$, $\left|\psi_-\right\rangle$, eqs.(4) or (5), gives the localized singly occupied dimer MO's $\left|\psi_A\right\rangle$, $\left|\psi_B\right\rangle$. $H_{AB}$ may then be expressed in terms of the monomer orbitals, where $\varepsilon_{A,B}^{mono} = \left\langle\psi_{A,B}^{mono}\middle|H\middle|\psi_{A,B}^{mono}\right\rangle$:

$$H_{AB} = \frac{1}{2}\left(\varepsilon_A^{mono}\left(N_+^2 - N_-^2\right) + \varepsilon_B^{mono}\left(N_+^2 - N_-^2\right) + 2\left\langle\psi_A^{mono}\middle|H\middle|\psi_B^{mono}\right\rangle\left(N_+^2 + N_-^2\right)\right) \qquad (8)$$

Assuming $S_{AB}^{mono}$ to be small the normalization factors can be approximated up to first order by $N_\pm^2 = \frac{1}{2}\left(1 \mp S_{AB}^{mono}\right)$ so that for a symmetric dimer eq.(8) simplifies to

$$H_{AB} = -\varepsilon^{mono} \cdot S_{AB}^{mono} + H_{AB}^{mono} \qquad (9)$$

where $\varepsilon^{mono} = \varepsilon_A^{mono} = \varepsilon_B^{mono} = \left\langle\psi_B^{mono}\middle|H\middle|\psi_B^{mono}\right\rangle$ and $H_{AB}^{mono} = \left\langle\psi_A^{mono}\middle|H\middle|\psi_B^{mono}\right\rangle$.

Alternatively, symmetrical orthogonalization of the monomer MO's yields localized orbitals, as well:

$$\left|\psi_A\right\rangle = N_A\left(\left|\psi_A^{mono}\right\rangle - \frac{1}{2}S_{AB}^{mono}\left|\psi_B^{mono}\right\rangle\right) \qquad (10)$$

with an analogous expression for $\left|\psi_B\right\rangle$ and the normalization factor $N_A = \left(1 - \frac{3}{4}\left(S_{AB}^{mono}\right)^2\right)^{-0.5}$. This construction should be more suitable for nonsymmetrical dimers since it circumvents the problem of numerically estimating $\gamma$ in eq.(4). With eq. (10) the transfer integral may be written as

$$H_{AB} = N_A \cdot N_B \cdot \left( H_{AB}^{mono} - \frac{1}{2} S_{AB}^{mono} \cdot \varepsilon_A^{mono} - \frac{1}{2} S_{AB}^{mono} \cdot \varepsilon_B^{mono} \right) \qquad (11)$$

where the small term proportional to $\left( S_{AB}^{mono} \right)^2$ is neglected. For the special case of a symmetric dimer eq.(11) reduces to eq.(9) to first order in $S_{AB}^{mono}$. In deriving an analytical expressions for the transfer integral the monomer approach will be discussed first. Since in some cases it has turned out that an analytical solution of the monomer problem is not possible with sufficient accuracy, alternatively it will be tried to directly construct the dimer-MO's $\left| \psi_+ \right\rangle$ and $\left| \psi_- \right\rangle$ what might still provide an analytical way for calculating the transfer integral. The different procedures for calculating $H_{AB}$ will be discussed on the example of planar doubly bridged $\left[ Cu_2 X_2 Y_4 \right]^{n-}$ dimers. Afterwards the analytical approaches will be applied to simple model complexes, namely singly and doubly bridged dimers, and the results will be compared with fully numerical calculations. Finally, the analytical transfer integrals will be utilized for reproducing and understanding the numerically calculated coupling constants of these complexes.

## III.   CALCULATION OF $H_{AB}$

### A .  Monomer approach

As already pointed out, the basic problem in calculating $H_{AB}$ analytically is the construction of appropriate singly occupied monomer MO's, eq.(7), with predominantly transition metal $d$-character. An analytical approach for constructing such orbitals has already been described [25] and turned out to be sufficiently accurate for describing $d$-orbital splitting pattern of TM-complexes [26]. This analytical approach transforms the full multi-centre MO-Hamiltonian of a TM surrounded by $N$ ligands into a single-centre problem in two steps: in the first step the Kohn-Sham equation is solved in linear combination of atomic orbital (LCAO) approximation with respect to metal atomic $d$-orbitals that are Schmidt-orthogonalized to the ligand atomic orbitals (AO's):

$$\left| \bar{\varphi}_d^X \right\rangle = N_d^X \left( \left| \varphi_d^X \right\rangle - \sum_i S_{di}^{Xx} \left| \varphi_i^x \right\rangle \right) \qquad (12)$$

where $S_{di}^{Xx}$ is the overlap matrix element between a metal $d$-orbital on centre $X$, $\left| \varphi_d^X \right\rangle$, and ligand AO $i$, $\left| \varphi_i^x \right\rangle$, on centre $x$ and $N_d^X = \left( 1 - \sum_i S_{id}^{xX} \right)^{-1/2}$ is the normalization factor. The sum runs over all ligand AO's which are assumed to be orthogonal to each other. In order to simplify the procedure the ligand AO's are combined linearly to obtain symmetry adapted group-orbitals corresponding to the symmetry of the singly occupied monomer MO. The summation in eq.(12) is then restricted to these group-orbitals. In the subsequent discussions ligand orbitals are always understood as group-orbitals, if not stated otherwise. Each AO is described by the product of a single Slater-type orbital (STO) and a real spherical harmonic, where the orbital exponent has turned out to be rather insensitive to geometrical changes of the dimers and is therefore taken as constant. In the second step, the Hamiltonian matrix calculated with the orthogonalized ligand and metal orbitals is diagonalized by a nondegenerate second-order perturbation calculation, i.e. the contribution of the nondiagonal elements to the energy of the ligand-orthogonalized $d$-orbital is accounted for in second order. If nondiagonal elements between ligand orbitals are assumed to be negligible, this matrix has nondiagonal elements only between the orthogonalized $d$-orbitals and the ligand group-orbitals. The complete analytical solution of the monomer problem requires the diagonal elements of the Hamiltonian matrix, the overlap-integrals between the STO's and the nondiagonal Hamiltonian matrix elements. The first ones are virtually constant with respect to geometrical changes of the dimers and are taken to be the orbital energies of the atoms in the molecule. The overlap integrals can easily be represented on closed form [27] whereas the nondiagonal elements are approximated by [25]:

$$H_{mn}^{XY} = \frac{1}{2} \left( H_{mm}^{XX} + H_{nn}^{YY} + 2\bar{\upsilon}_{mn}^{XY} \right) \cdot S_{mn}^{XY} \qquad (13)$$

where $m,n$ are any orbitals on the centres $X,Y$. $H_{mm}^{XX}$ and $H_{nn}^{YY}$ are the diagonal elements of the Hamiltonian matrix approximated by the orbital energies of atoms in the molecule. $\bar{\upsilon}_{mn}^{XY}$ is an angular-independent averaged potential of 2- and 3-centre Coulomb and exchange integrals $\upsilon_2, \bar{\upsilon}_3, \bar{\upsilon}_{xc}$ that are in good approximation proportional to $S_{mn}^{XY}$ [25]:

$$\overline{\upsilon}_{mn}^{XY} = \upsilon_{mn}^{XY} / S_{mn}^{XY} = \left( \upsilon_2 + \overline{\upsilon}_3 + \overline{\upsilon}_{xc} \right) / S_{mn}^{XY} \tag{14}$$

$\overline{\upsilon}_{mn}^{XY}$ is always negative and proportional to $1/R^2$ where $R$ is the bonding distance between $X$ and $Y$ [25].

The resulting analytical approach, denoted as $d$-Hamiltonian, has been proven to be sufficiently accurate for describing the antibonding MO's with predominantly $d$-character of transition metal monomers [25,26] representing therefore an adequate starting point for calculating $H_{AB}$. In order to derive sufficiently simple analytical formulas, ligand-ligand interactions as well as metal $4s$- and $4p$-orbitals were neglected in the monomer approach. Both are assumed to have negligible effects on the magnetic coupling mechanism in ionic compound, as well. Decomposing the dimer into two monomers $A$ and $B$ the energy of the singly occupied monomer MO at centre $A$ is given as [25]

$$\varepsilon_A^{mono} = H_{dd}^{AA} - \sum_i \left( 2\overline{\upsilon}_{id}^{aA} + \frac{\left( \Delta_{id}^{aA} - 2\overline{\upsilon}_{id}^{aA} \right)^2}{4\Delta_{id}^{aA}} \right) \cdot S_{id}^{aA,2} \tag{15}$$

with $\Delta_{id}^{aA} = H_{ii}^{aa} - H_{dd}^{AA}$ and a similar expression for monomer $B$. The corresponding eigenfunction to $\varepsilon_A^{mono}$, i.e. the monomer MO, is given to first order in $S_{id}^{aA}$ by

$$\psi_A^{mono} = N_A^{mono} \left( \left| \varphi_d^A \right\rangle - \sum_i \tau_{id}^{aA} S_{id}^{aA} \cdot \left| \varphi_i^a \right\rangle \right) \tag{16}$$

where the coefficient $\tau_{id}^{aA} = \dfrac{\Delta_{id}^{aA} + 2\overline{\upsilon}_{id}^{aA}}{2\Delta_{id}^{aA}}$ is angular-independent and contains quantities that can easily be derived e.g. from spectroscopic data. Consequently, the whole angular dependence is contained in the overlap integrals $S_{id}^{aA}$. Analogous expressions are obtained for the monomer $B$, with $\varphi_{d'}^B$ and $\varphi_i^b$. The overlap integral between these monomer MO's at centres $A$ and $B$ is derived as

$$\begin{aligned} S_{AB}^{mono} &= \left\langle \psi_A^{mono} \middle| \psi_B^{mono} \right\rangle \\ &= N_A^{mono} N_B^{mono} \cdot \left( S_{dd'}^{AB} - \sum_i \tau_{d'i}^{Bb} \cdot S_{id'}^{bB} \cdot S_{di}^{Ab} - \sum_i \tau_{di}^{Aa} \cdot S_{id}^{aA} \cdot S_{d'i}^{Ba} + \sum_{i,j} \tau_{di}^{Aa} \tau_{d'i}^{Bb} \cdot S_{id}^{aA} S_{id'}^{bB} \cdot S_{ij}^{ab} \right) \end{aligned} \tag{17}$$

where $i$, $j$ run over all ligand orbitals on the centres $A$ and $B$, respectively. In order to simplify this result, the following assumptions are made:

(i)     ligand-ligand interactions are assumed to be negligible

$$H_{ij}^{aa} = H_{ij}^{aa} \cdot \delta_{kl}; \quad H_{ij}^{bb} = H_{ij}^{bb} \cdot \delta_{ij}; \quad H_{kl}^{ab} = H_{kl}^{ba} = H_{kl}^{ab} \cdot \delta_{kl}$$

$$S_{ij}^{aa} = S_{ij}^{bb} = \delta_{ij}; \quad S_{kl}^{ab} = S_{kl}^{ba} = S_{kl}^{ab} \cdot \delta_{kl}$$

where $i,j$ run over all ligand orbitals whereas $k,l$ denote bridging orbitals

(ii)     interactions between the metal and the terminal ligands at different monomers are neglected

(iii)     for the special case of a symmetric complex further simplifications result

$$H_{dd}^{AA} = H_{d'd'}^{BB} = H_{dd} \qquad H_{ii}^{aa} = H_{ii}^{bb} = H_{ii} \qquad N_A^{mono} = N_B^{mono} = N^{mono}$$

$$\tau_{di}^{Aa} = \tau_{d'i}^{Bb} = \tau_{di}; \quad \tau_{dk}^{Ab} = \tau_{d'k}^{Bb} = \tau_{dk} \qquad S_{dk}^{Aa} = S_{d'k}^{Bb} = (\pm 1) S_{dk}^{Ab} = (\pm 1) S_{d'k}^{Ba} \tag{18}$$

Thus, the site indices $A,B,a,b$ can be omitted in these quantities. The factors $(\pm 1)$ are required since the bridging group-orbitals for the two monomers may differ by sign, e.g. for a planar doubly bridged $\left[ Cu_2 X_2 Y_4 \right]^{n-}$ complex where the $xz$-plane is the molecular plane and $z$ is the internuclear axis (figure 1A) the following relations are fulfilled:

$$\left| \varphi_s^a \right\rangle = -\left| \varphi_s^b \right\rangle; \quad \left| \varphi_z^a \right\rangle = \left| \varphi_z^b \right\rangle; \quad \left| \varphi_x^a \right\rangle = -\left| \varphi_x^b \right\rangle \tag{19}$$

where $s, z, x$ denote $s$, $p_z$ and $p_x$ group orbitals. With these assumptions eq.(17) reduces to

$$S_{AB}^{mono} = \left( N^{mono} \right)^2 \cdot \left( S_{dd'}^{AB} - \sum_k \left( 2\tau_{dk} - \tau_{dk}^2 \right) \cdot S_{kd}^2 \cdot (\pm 1) \right) \tag{20}$$

where the sign of $(\pm 1)$ depends on the bridging orbital $k$. Since $S_{AB}^{mono} < 0.05$, as will be shown below (table I), terms of order $\left( S_{AB}^{mono} \right)^n$ with $(n > 1)$ can be neglected. Inserting eq.(20) together with eq.(10) into eq.(6) and neglecting higher order terms in $\left( S_{AB}^{mono} \right)^n$ and $S_{dk}^{2n}$ $(n > 1)$ yields [28]

$$H_{AB} = \overline{\upsilon}_{dd'}^{AB} \cdot S_{dd'}^{AB} - \sum_k \frac{\left(\Delta_{kd} + 2\overline{\upsilon}_{kd}\right)^2}{4\Delta_{kd}} \cdot S_{dk}^2 \cdot \left(\pm 1\right) \qquad (21)$$

The first term, describing the direct $d$-$d$ interaction, stems from the approximation for the nondiagonal element $H_{dd'}^{AB}$ between atomic $d$-orbitals on different centres (cf. eq.(13)). For ionic compounds as fluorides (see below) and oxides this simple expression for the transfer integral supplies results that can easily be interpreted and are in reasonable agreement with fully numerical calculations. Moreover, eq.(21) can directly be improved by including further interactions.

However, if the energy differences between metal and ligand orbitals decrease or the nondiagonal elements are large, the perturbation calculation generally leads to poor results. In this case an alternative approach for the diagonalization will be used. The small contributions of the ligand $s$-orbitals can still be included as perturbations but will at first be neglected in the following discussion of the diagonalization procedure. The diagonalization of the Hamiltonian matrix is carried out in two different ways:

**Bridging ligand only method.** Assuming that in a symmetric complex the contributions from the terminal ligands cancel each other in the difference $\varepsilon_+ - \varepsilon_-$, eq.(6), only the bridging ligands are explicitly included in the diagonalization procedure. The $m$ diagonal-elements of the $p$-orbitals of the bridging ligands are averaged: $\overline{H}_{pp} = \frac{1}{m}\sum_p H_{pp}$. This approximation is justified if the $p$-orbital energies are similar and also energetically well separated from the orthogonalized metal $d$-orbital. The energy of the magnetic monomer orbital at site $A$ and the corresponding eigenvector $\left|\psi_A^{mono}\right\rangle$ are derived as

$$\varepsilon_A^{mono} = \frac{1}{2}\left(\tilde{H}_{dd}^{AA} + \overline{H}_{pp}^{aa} + \sqrt{\left(\tilde{\Delta}_{pd}^{aA}\right)^2 + 4\sum_p\left(\tilde{H}_{pd}^{aA}\right)^2}\right) \qquad (22)$$

$$\left|\psi_A^{mono}\right\rangle = N_A^{mono}\left(\tilde{\alpha}_d^A \cdot \left|\tilde{\varphi}_d^A\right\rangle + \sum_p \beta_p^a \left|\varphi_p^a\right\rangle\right) \qquad (23)$$

where $\tilde{\Delta}_{pd}^{aA} = \bar{H}_{pp}^{aa} - \tilde{H}_{dd}^{AA}$. $p$ runs over all bridging $p$-orbitals. $\tilde{H}_{dd}^{AA} = \langle \tilde{\varphi}_d^A | H | \tilde{\varphi}_d^A \rangle$ is the energy of the ligand-orthogonalized $d$-orbital (eq.(12)) and $\tilde{H}_{id}^{aA} = \langle \tilde{\varphi}_d^A | H | \varphi_i^a \rangle$ is the nondiagonal element with a ligand group orbital $i$. The coefficients are obtained as

$$\tilde{\alpha}_d^A = \frac{1}{2 \cdot \tilde{H}_{md}^{aA}} \left( -\tilde{\Delta}_{pd}^{aA} + \sqrt{\left(\tilde{\Delta}_{pd}^{aA}\right)^2 + 4 \cdot \sum_p \left(\tilde{H}_{pd}^{aA}\right)^2} \right); \quad \beta_p^a = \tilde{H}_{pd}^{aA} / \tilde{H}_{md}^{aA}; \quad \beta_m^a = \tilde{H}_{md}^{aA} / \tilde{H}_{md}^{aA} = 1$$

where $m$ is one of the $p$-orbitals. Rewriting the eigenvector (23) in terms of the atomic $d$-orbital $|\varphi_d^A\rangle$ gives

$$|\psi_A^{mono}\rangle = N_A^{mono} \left( \alpha_d^A \cdot |\varphi_d^A\rangle + \sum_s \gamma_s^a \cdot |\varphi_s^a\rangle + \sum_p \gamma_p^a |\varphi_p^a\rangle \right) = N_A^{mono} \left( \alpha_d^A |\varphi_d^A\rangle + \sum_i \gamma_i^a |\varphi_i^a\rangle \right) \quad (24)$$

where $\alpha_d^A = \tilde{\alpha}_d^A \cdot N_d^A$; $\gamma_s^a = -\tilde{\alpha}_d^A N_d^A \cdot S_{sd}^{aA}$; $\gamma_p^a = \beta_p^a - \tilde{\alpha}_d^A N_d^A \cdot S_{pd}^{aA}$. The contribution from the $s$-orbital arises from the orthogonalization. Its small contribution to the diagonalization may be considered via adding $-\left(\tilde{H}_{sd}^{aA}\right)^2 \big/ \Delta_{sd}^{aA}$ to eq.(24) and accordingly $-\tilde{H}_{sd}^{aA} \big/ \Delta_{sd}^{aA} \cdot \tilde{\alpha}_d^A$ to $\gamma_s^a$. The contributions of the terminal ligands, arising solely from the orthogonalization, are given as $\gamma_p^a = -\tilde{\alpha}_d \cdot N_d \cdot S_{pd}^{aA}$. Therefore, $i$ in eq.(24) runs over all ligand orbitals on the respective monomer.

**Averaging over all ligands.** If the terminal ligands of the monomers $A$ and $B$ are different no cancellation of their contributions to the transfer integral can be expected. Therefore, it may be tried to extend the averaging over all ligand $p$-orbitals. Eqs.(22)-(24) can be applied in this case, as well, if the summations include also the terminal ligands. This approximation requires again a clear energetic separation of metal and ligand orbitals.

The nondiagonal element $H_{AB}^{mono} = \langle \psi_A^{mono} | H | \psi_B^{mono} \rangle$ as well as the overlap integral $S_{AB}^{mono} = \langle \psi_A^{mono} | \psi_B^{mono} \rangle$ can directly be calculated with either method. Using the relations (18) and $\alpha_d^A = \alpha_{d'}^B = \alpha$ for a symmetric complex, these integrals are

$$H_{AB}^{mono} = \left\langle \psi_A^{mono} \middle| H \middle| \psi_B^{mono} \right\rangle = \left(N^{mono}\right)^2 \left( \alpha^2 \cdot H_{dd'}^{AB} + \sum_k \left(2\alpha \cdot \gamma_k \cdot H_{dk} + \gamma_k^2 \cdot H_{kk}\right)(\pm 1) \right) \qquad (25)$$

$$S_{AB}^{mono} = \left\langle \psi_A^{mono} \middle| \psi_B^{mono} \right\rangle = \left(N^{mono}\right)^2 \left( \alpha^2 \cdot S_{dd'}^{AB} + \sum_k \left(2\alpha \cdot \gamma_k \cdot S_{kd} + \gamma_k^2\right)(\pm 1) \right) \qquad (26)$$

and $\varepsilon_A^{mono}$ is obtained as

$$\varepsilon_A^{mono} = \left\langle \psi_A^{mono} \middle| H \middle| \psi_A^{mono} \right\rangle = \left(N_A^{mono}\right)^2 \left( \alpha^2 H_{dd}^{AA} + \sum_i \left(2a\gamma_i H_{id} + \gamma_i^2 H_{ii}\right) \right) \qquad (27)$$

with $i$ running over all ligands on centre $A$. Inserting eqs.(25)-(27) into eq.(9) and neglecting terms proportional to $\left(S_{du}\right)^n$ with $n > 2$ yields

$$H_{AB} = -\alpha^4 H_{dd}^{AA} \cdot S_{dd'}^{AB} + \alpha^2 H_{dd'}^{AB} - \sum_k \gamma_k \cdot \left( \alpha^2 H_{dd}^{AA} \cdot \left(2\alpha S_{dk} + \gamma_k\right) - 2H_{dk}\alpha - H_{kk}\gamma_k \right) \cdot (\pm 1) \qquad (28)$$

This procedure avoids the problems of a perturbation calculation and therefore might supply better results than eq.(21). In the case of nonnegligible interactions within the terminal or bridging ligands it might be necessary to orthogonalize the ligand orbitals among each other and include also their nondiagonal Hamiltonian matrix elements. This is expected to be required for ligands of the third or higher period or for molecular ligands where the appropriate MO's interacting with the $d$-orbitals have to be constructed. In summary, the monomer approach presents the simplest way to calculate directly the transfer integral $H_{AB}$. Moreover, it enables decomposition into the single orbital contributions.

In this work as first step the applicability of such an analytical approach is investigated on the simplest model systems, which are symmetrical. An extension of the monomer approach to unsymmetrical and heteronuclear compounds is possible without difficulties and results are going to be published soon. Moreover, the monomer approach has turned out to work in principle also for strongly covalent complexes containing e.g. chlorine or sulphur ligands. However, the strong interactions of the ligand orbitals may cause a significant energy difference of all the ligand $p$-orbitals so that no averaging and therefore no simple analytical diagonalization is possible. This problem may be bypassed by directly constructing the dimer MO's $|\psi_+\rangle$ and $|\psi_-\rangle$.

**B. Dimer approach**

The direct solution of the dimer problem, is analogous to the formalism developed for the monomers. The two dimer MO's belong to different irreducible representations of the symmetry group of the whole molecule. Thus, group-orbitals encompassing now orbitals from both centres $A$ and $B$ have to be constructed according to the respective symmetry (figure 1B). With regard to this basis the Hamiltonian matrix splits into blocked matrices, which can be diagonalized separately by the methods discussed for the monomer approach. The respective highest eigenvalues (lowest binding energy) correspond to the energy of the magnetic orbitals $\varepsilon_+$, $\varepsilon_-$, see eq.(6), and diagonalization yields the eigenvalues

$$\varepsilon_\pm = \frac{1}{2}\left( \tilde{H}_{dd}^\pm + \overline{H}_{pp}^\pm + \sqrt{\left(\tilde{\Delta}_{pd}^\pm\right)^2 + 4\cdot\sum_p \left(\tilde{H}_{pd}^\pm\right)^2} \right) \tag{29}$$

where $\tilde{\Delta}_{pd}^\pm = \overline{H}_{pp}^\pm - \tilde{H}_{dd}^\pm$. The $s$-orbitals may be considered again as perturbations, the ligand-ligand interactions may be included in the same manner as described above. For an ionic system, as the Cu fluorides, it is also possible to derive $\varepsilon_\pm$ from a perturbation calculation, with the simple result

$$\varepsilon_\pm = \tilde{H}_{dd}^\pm - \sum_i \frac{\left(\tilde{H}_{id}^\pm\right)^2}{\Delta_{id}^\pm} \tag{30}$$

## IV. APPLICATION

In a first series of applications the transfer integral is derived and analyzed for singly and doubly bridged Cu-F dimers. The results of the analytical approach are compared with fully numerical calculations in the local density approximation by the spin-polarized self-consistent charge $X\alpha$ (SCC-$X\alpha$) method [29,30], $H_{AB}$ corresponds to half of the energy difference of the HOMO and LUMO orbitals in the spin restricted calculation, cf. eq.(6).

## A. The monomer approach for [Cu$_2$F$_6$]$^{2-}$

This ionic model dimer of symmetry $D_{2h}$ is assumed to have a Cu-F distance of 1.94Å and an angle of 93° between Cu and the terminal ligands. The transfer integral will be calculated for bridging angles $\theta$ between 82.5° and 125°.

The two CuF$_4$-monomers exhibit symmetry $C_{2v}$ (figure 2) with magnetic orbitals transforming after the irreducible representation $b_1$ with predominantly Cu($d_{xz}$)-character if the $xz$-plane is defined as the molecular plane. According to the antibonding character of this orbital the metal ligand overlap matrix element is negative.

**FIG 2.**

The analytical calculations are carried out in three different ways, viz. according to eq.(21), denoted as mon1, and the two approaches for direct diagonalization, i.e. the bridging ligand only method (mon2) as well as the method averaging over all ligand energies (mon3) both using eq.(28). The energy difference between the $p$-orbitals of terminal and bridging ligands before averaging is about 2.5eV and the minimum difference to the Cu(3$d$)-orbital about 3.5eV. The resulting magnetic monomer orbitals exhibit large contributions (>0.90) from Cu(3$d$), contributions in the range 0.2-0.3 from the ligand $p$-orbitals and also small but nonnegligible ones from the ligand $s$-orbitals (0.05-0.10). Thus the spins are well localized at the transition metal.

The key quantities in eq.(25)-(27) required for the calculation of $H_{AB}$, eq.(9), obtained with mon2 and mon3 are given in table I for bridging angles of 82.5°, 90° and 110°.

**TABLE I.**

The data merely roughly confirm the empirically expected correlation between the overlap integral between the monomer orbitals $S_{AB}^{mono}$ and the transfer integral $H_{AB}$ since $H_{AB}^{mono}$, eq.(9), is not exactly proportional to $S_{AB}^{mono}$. Such a proportionality is obtained if only *p-d* interactions are considered. Therefore, a more detailed physical interpretation may be supplied by analyzing separately the single orbital interactions given by eq.(21) for mon1 and eq.(28) for mon2 (cf. table II). The application of eq.(28) with mon3 yields very similar results for this dimer and is therefore omitted.

**TABLE II.**

These results confirm the empirical Goodenough-Kanamori rules [4,5] predicting the compensation of *p-d* interactions at 90°. However, due to direct *d-d* interactions and *s-d* contributions that are indeed not negligible, a vanishing transfer-integral is not obtained for an exact rectangular arrangement, but the zero is shifted to larger angles (figure 3).

Next, the transfer integral $H_{AB}$ for a series of different bridging angles $\theta$, calculated with the three monomer methods (mon1-mon3) and eq.(28) for mon2, are compared with the numerical values from SCC-X$\alpha$. The results are displayed in figure 3. The differences between the columns of mon2 and eq.(28) are due to small simplifications arising from expanding eq.(9) to yield eq.(28) (see above). As shown in figure 3, the numerical $H_{AB}$ is an approximately linear function of $\theta$ vanishing at about 92°. The various analytical calculations exhibit a uniform shift of the zero up to about 96°. Model calculations have shown that this is due to the neglect of ligand-ligand interactions. The slopes agree quite well with the numerical curve, except mon3 exhibiting an increasing slope above 110° instead of a flattening. Especially with respect to the simplifying model assumptions, this must be considered as a satisfying result confirming the suitability of the analytical approach for investigating magneto-structural correlations.

**FIG 3.**

**B. The dimer approach for a $[Cu_2F_6]^{2-}$ complex**

The energies of the two magnetic dimer orbitals are calculated according to eq.(29). Again, the diagonalization is performed by averaging over all ligands (dim1) as well as with the bridging ligand only method (dim2) where also eq.(30) is used. The results are presented in figure 4.

**FIG 4.**

Again, analytical and numerical results for $H_{AB}$ are very similar with each other. Consequently, this analysis demonstrates that at least for these symmetric double bridged dimers, both the monomer and the dimer approaches represent appropriate starting points for the calculation of $H_{AB}$ and its geometrical dependencies. The larger slope of the transfer integral calculated with eq.(30) arises from the different size of the matrix elements $\tilde{H}_{pd}^{+}$ and $\tilde{H}_{pd}^{-}$ of the symmetric and the unsymmetrical magnetic orbital, respectively leading to a differently well fulfilled applicability of the perturbation calculation. The contributions from orthogonalization and diagonalization for dim2 are given in table III.

**TABLE III.**

At small and large angles the main contribution arises from diagonalization, while around 90-95° the contributions become comparable and compensate near 96°. Hence, too simple approximations for the diagonalization procedure lead to worse results.

**C. Variation of the bonding distance in $[Cu_2F_6]^{2-}$**

In addition to the angular dependence of the transfer integral, the distance dependence is investigated for the doubly bridged complex with fixed bridging angle of 90°. In this case, the empirical rules for superexchange [4,5] predict ferromagnetic coupling. The bonding distance $d_{Cu-F}$ between Cu and the bridging ligands is varied between 1.7 and 2.2Å whereas the distance to the terminal ligands is kept constant. At small distances the ligand-ligand interactions, even between terminal and bridging ligands are expected to be nonnegligible so that the dimer approach should be more suitable. The interactions between the bridging orbitals as well as between bridging and terminal $p_z$ orbitals are included. For diagonalization the bridging ligand only method, dim2, is being used since the $p$-orbital energies between terminal and bridging ligands may differ by several eV for small $d_{Cu-F}$ so that averaging is inappropriate. The results are shown in figure 5 together with the values obtained when ligand-ligand interactions are neglected, as well as a diagonalization using perturbation calculation, eq.(30).

**FIG 5.**

As expected, ligand-ligand interactions are of crucial importance for the small bonding distances. Unfortunately, due to these interactions the contributions from the terminal ligands are somewhat different for the symmetric and the antisymmetric magnetic orbital so that the compensation is not complete leading to significant deviations between dim2 and the numerical values at small bonding distances. However, since the ligand and metal orbitals are separated energetically by more than 4eV a perturbation calculation is feasible for diagonalization. The results are depicted in figure 5, denoted as eq.(30), and are in reasonable accordance with the fully numerical calculation. The results of the monomer approaches without ligand-ligand interactions are shown in figure 6.

**FIG 6.**

As in the dim2* calculation, the transfer integral is strongly overestimated for small bonding distances though qualitatively the correct behaviour of the transfer integral is reproduced. In order to understand the reasons for the deviations from the 90° rule of ferromagnetic coupling, the contributions from the direct *d-d* and *s-d* interactions have to be analyzed (cf. table IV). Since the values are taken from the mon1 calculations the *p-d* contributions cancel each other exactly. Consequently, the large value of the transfer integral for small bridging angles arises from the interaction with the ligand s-orbitals.

**TABLE IV.**

### D. Singly bridged [Cu$_2$F$_7$]$^{3-}$

This complex is treated only on the basis of the monomer approaches. Each monomer has symmetry $C_{2v}$ with an assumed bonding distance of 1.94Å. Unlike the planar, doubly bridged dimers where the metal *d*-orbital contributing to the magnetic monomer orbital was a pure $d_{xz}$-orbital, for nonlinear bridges (figure 7) it is now a linear combination of three real *d*-orbitals of different *m* quantum numbers. Since the coefficients of the linear combination are a priori not known, the *d*-orbitals to different *m* have to be separately orthogonalized to the ligands. After doing so, the *d*-orbitals are no longer orthogonal to each other, so that a second orthogonalization step is required. The resulting Hamiltonian matrix has nondiagonal elements in more than one column and row preventing an analytical diagonalization. Therefore, the linear combination should be estimated first: When both monomers are placed in the *xy*-plane, i.e. the dimer is planar, the *d*-orbital contributing to the singly occupied monomer MO's is a pure $d_{x^2-y^2}$. The small contribution from $d_{z^2}$ arising from the slight distortion of the $D_{4h}$ monomer-symmetry in the dimer may safely be ignored. Assuming that the two magnetic centers interact only via their bridging ligand, the linear combination of the *d*-orbitals is in good approximation given by rotating the $d_{x^2-y^2}$-orbital into the monomer plane:

$$\left| d_{rot} \right\rangle = \left( \cos\alpha \cdot \sin\alpha \right) \cdot \left| d_{yz} \right\rangle + \left( \frac{1}{2}\sqrt{3}\sin^2\alpha \right) \cdot \left| d_{z^2} \right\rangle + \left( 1 - \frac{1}{2}\sin^2\alpha \right) \cdot \left| d_{x^2-y^2} \right\rangle \qquad (31)$$

where $\alpha[°] = (180 - \theta)/2$. Using eq.(31) for the magnetic *d*-orbital allows treating this singly bridged dimer in the same manner as the doubly bridged one.

**FIG 7.**

The analytical calculations are performed by the three monomer approaches mon1-mon3. The required quantities, eqs.(25)-(27), are given in table V for three bridging angles.

**TABLE V.**

Again, there is only a rough correlation between $H_{AB}$ and $S_{AB}^{mono}$ as discussed previously for the doubly bridged dimer. Unlike this doubly bridged complex the slope of the transfer integral with regard to increasing bridging angles is negative (figure 8) due to the mirror plane ($xz$) between the monomers corresponding to the relations

$$\left| \varphi_s^a \right\rangle = \left| \varphi_s^b \right\rangle ; \quad \left| \varphi_y^a \right\rangle = -\left| \varphi_y^b \right\rangle ; \quad \left| \varphi_z^a \right\rangle = \left| \varphi_z^b \right\rangle \tag{32}$$

Accordingly, the overlap between the monomers via the bridging orbital $\left| \varphi_y^a \right\rangle$ parallel to the internuclear axis is negative, whereas for the doubly bridged complex this interaction, via $\left| \varphi_z \right\rangle$, is positive, eq.(19). A similar behaviour is obtained for the other orbitals. As shown in figure 8 analytical and fully numerical results compare very well.

**FIG 8.**

**V. CORRELATION BETWEEN $H_{AB}$ AND $J$**

Assuming that the variations with respect to geometry of both the ferromagnetic term, $K_{AB}$, and the effective Hubbard $U$ with $U = U_{AA} - U_{AB}$, in eq.(2) are small, the behaviour of $J$ as function of the geometry is exclusively determined by the variations of $H_{AB}$. Accordingly, in exploring the geometry dependence the coupling constant $J$ can be transformed into a parameterized form

$$J\left(fit\right) = C - 2 \cdot H_{AB}^2 \cdot f \qquad (33)$$

where the parameters $C$ and $f$ are positive, geometry-independent constants. The applicability of eq.(33) is tested on several different Cu-F dimers, where both numerical and analytical transfer integrals are used. The results are compared with fully numerical calculations of the coupling constant $J$ using the broken symmetry (bs) formalism [8]. In addition to the SCC-X$\alpha$ method, the full potential local orbital (FPLO) code fplo7.00-28 [31] is applied where the nonrelativistic mode and an exchange-only exchange-correlation potential has been used. Additionally, no Madelung potential is used in the FPLO-calculations.

## A. Doubly bridged [Cu$_2$F$_6$]$^{2-}$

In a first attempt the two-electron integrals in eq.(2) are evaluated numerically for the $[Cu_2F_6]^{2-}$ complex, using localized magnetic dimer orbitals constructed from the Kohn-Sham orbitals from the SCC-X$\alpha$ calculations. The numerical calculations were performed with *Mathematica7®* enabling a direct calculation with the localized magnetic dimer-orbitals. Alternatively, Fortran programs were developed. The results for $K_{AB}$ and $U$ as a function of the bridging angle vary within 10% of their absolute values. Thus, $H_{AB}$ is indeed the crucial parameter for exploring magneto-structural correlations. However, the numerical values of the two-electron integrals, especially $U$ with about 9eV, are considerably too large. This is well known and may be traced back to the neglect of screening and rearrangement effects. On the other hand the qualitative trend of $U$ and $K_{AB}$ being constant should not be affected, so that eq.(33) is a good approximation.

In the next step, the geometrical variation of the coupling constant $J$ is compared for:

(i)     $J$(bs), from the broken symmetry calculations [8]

$$J\left(bs\right) = \frac{E\left(S_{\max}\right) - E\left(S_{\min}\right)}{S_{\max}^2 - S_{\min}^2} \qquad (34)$$

where $S_{max}$ and $S_{min}$ are the spins of the high-spin and low-spin states, respectively, of the dimer, and $E(S)$ is the numerically obtained total energy of the spin state $S$.

(ii)    $J$(fit), obtained by scaling the numerical $H_{AB'}$ according to eq.(33)

(iii)   $J(analyt)$, calculated with eq.(33) and $H_{AB}$ derived from the various analytical approches

Choosing $C = 80$ and $f = 1/13000$ in eq.(33) for scaling the numerical $H_{AB}$ from the SCC-X$\alpha$ calculation (cf. figure 3 or 4) yields coupling constants $J(fit)$ in very good agreement with $J(bs)$ as calculated by the SCC-X$\alpha$ method using the broken symmetry formalism (figure 9) merely at larger angles some deviations occur.

**FIG 9.**

In order to assess these results and the validity of the assumptions analogous calculations were performed by FPLO.

**TABLE VI.**

As shown in table VI and figure 9, calculated and fitted coupling constants are virtually identical over the whole range of $\theta$, again confirming that magneto-structural correlations are determined by the variation of the transfer integral $H_{AB}$, indeed. The differences between SCC-X$\alpha$ and FPLO may be traced back basically to the different basis sets [29,31].

Applying eq.(33) to the various monomer and dimer approaches (cf. figures 3 and 4) yields the results displayed in figure 10.

**FIG 10.**

With regard to the simplifying model assumptions the analytical approaches reproduce the qualitative trend satisfactorily, supplying an easy way for interpreting the magnetic behaviour.

**B. [Cu$_2$F$_6$]$^{2-}$ with varying bonding distance**

The behaviour of the coupling constant as a function of the bonding distance between Cu and the bridging ligands, $d_{Cu-F}$, calculated with SCC-X$\alpha$ can be well reproduced using eq.(33) and the numerical transfer integrals (figure 5 or 6). The coupling constant obtained with FPLO is again qualitatively the same as for SCC-X$\alpha$. The shift of the ferromagnetic region to smaller bonding distances is an effect of the missing Madelung potential in the FPLO calculation. The results of both numerical methods are shown in figure 11.

**FIG 11.**

Again, there are small deviations between $J(bs)$ and $J(fit)$ in the SCC-X$\alpha$ results which do not occur in the FPLO calculations. A very similar agreement of $J(bs)$ and $J(analyt)$ as for the numerical transfer integral from SCC-X$\alpha$ can be obtained with the analytical $H_{AB}$ (cf. figure 5). The results are depicted in figure 12.

**FIG 12.**

**C. Singly bridged [Cu$_2$F$_7$]$^{3-}$.**

Finally, analogous calculations have been carried out for the singly bridged complex $[Cu_2F_7]^{3-}$. The results from SCC-X$\alpha$ and FPLO are summarized in figure 13.

**FIG 13.**

The $J(analyt)$ calculated with the analytical transfer integrals depicted in figure 8 are shown in figure 14. The agreement between the broken symmetry and the fitted results is again satisfactorily even for the analytical transfer integrals. However, for small bridging angles, the numerical calculation predicts a positive coupling constant while from the fitting procedure a negative $J$ is obtained. This discrepancy might be attributed to nonconstant two-electron integrals or additional higher order contributions. By contrast, this problem is not present in FPLO (figure 13). Most likely the discrepancy between $J(bs)$ and $J(fit)$ in SCC-X$\alpha$ is an artefact of the simple basis set used. The good agreement of the coupling constants of both numerical codes calculated with the broken symmetry formalism confirms that most of the errors done in the calculation of the total energies of the two magnetic states cancel each other in the energy difference, eq.(34).

**FIG 14.**

## VI. CONCLUSION

Analytical approaches, viz. various monomer and dimer approximations, were developed that enable the analytical calculation of transfer integrals $H_{AB}$ and an estimate of the superexchange coupling constant. The monomer approaches supply compact and simple formulas especially suitable for ionic compounds, whereas the dimer approach additionally allows dealing with compounds exhibiting strong orbital interactions between transition metals and ligands. With regard to the simplifying model assumptions, the analytical results are in satisfactory agreement with fully numerical calculations on Cu-F model dimers. As commonly assumed and confirmed by rough numerical estimates, the ferromagnetic contribution $K_{AB}$ and the effective electrostatic on-site interaction $U$ contained in the formula of the superexchange coupling constant can be taken as approximately constant. Since both cannot directly be calculated with sufficient reliability, two constant parameters were introduced. The applicability of the resulting parameterized formula for the superexchange coupling constant was tested by fitting numerically determined $H_{AB}$ to numerical coupling constants calculated with two different numerical methods. Especially the results from the numerical calculations with the FPLO code confirm that this is an excellent approximation. Therefore, the magneto-structural correlations in the superexchange coupled model compounds with fluorine ligands are fully described by the transfer integral. Similar results are obtained for copper-complexes with sulphur, chlorine or OH-ligands, as well, strongly interacting with the Cu-orbitals. The results for these covalent complexes will be published elsewhere. Consequently, if the magnetic behaviour of a specific complex cannot be described as a function of the square of the transfer integral, i.e. with the suggested parameterized formula, this can be regarded as a hint that higher order terms, as ferromagnetic kinetic exchange, come into play. Such additional contributions occur if e.g. if the magnetic orbitals come close to fully occupied orbitals due to structural changes. Thus, the present approach may also be used as indicating such contributions.

As the main advantage, compared with other methods, these analytical approaches provide a simple scheme for estimating magnetic coupling constants on the basis of orbital interactions beyond simple empirical rules but without performing fully numerical calculations from which the crucial interactions are usually difficult to detect. Finally, it has to be emphasized that the applicability of this approach is not limited to symmetric dimers with one magnetic orbital per metal centre. Recent studies have shown that it is applicable also to compounds with more than one unpaired electron per metal site (e.g. Fe-complexes). Moreover, especially the monomer approach can be extended to heteronuclear complexes. These results will be published soon.

**ACKNOWLEDGEMENTS**

This work has been financially supported by the Austrian Fonds zur Förderung der wissenschaftlichen Forschung (Project-No. P20503).

**TABLES**

Table 1. Quantities required for calculating $H_{AB}$ with eq.(9) for three different bridging angles with the methods mon2, mon3.

|  | 82.5° | | 90° | | 110° | |
|---|---|---|---|---|---|---|
|  | mon2 | mon3 | mon2 | mon3 | mon2 | mon3 |
| $\varepsilon_A^{mono}$ (eV) | -11.115 | -10.541 | -11.060 | -10.500 | -11.133 | -10.583 |
| $H_{AB}^{mono}$ (eV) | -0.111 | -0.060 | -0.135 | -0.122 | -0.003 | -0.095 |
| $S_{AB}^{mono}$ | -0.014 | -0.016 | -0.0002 | -0.0002 | 0.023 | 0.029 |
| $H_{AB}$ (cm$^{-1}$) | -2156 | -1865 | -1108 | -1004 | 2066 | 1720 |

Table 2. Contributions to $H_{AB}$ (in cm$^{-1}$) from eqs.(21) and (28) for different bridging angles. *d-d*: direct

interaction between metals. *k-d*: interaction between metal *d* and a bridging orbital *k*.

| method | $\theta$ [°] | *d-d* | *s-d* | *z-d* | *x-d* | $H_{AB}$ |
|--------|--------|-------|-------|-------|-------|----------|
| eq.(21) | 82.5 | 980 | -1627 | 2814 | -4627 | -2460 |
| | 90 | 514 | -1643 | 3717 | -3717 | -1129 |
| | 110 | 102 | -1368 | 5051 | -1284 | 2501 |
| eq.(28) | 82.5 | 977 | -1580 | 2637 | -4307 | -2282 |
| | 90 | 513 | -1615 | 3486 | -3486 | -1089 |
| | 110 | 101 | -1354 | 4799 | -1219 | 2328 |

Table 3. Contributions from orthogonalization (orth.) and diagonalization (dia) for dim2 (in cm$^{-1}$).

| $\theta$ [°] | orth. | dia. | $H_{AB}$ |
|---|---|---|---|
| 82.5 | -793 | -1554 | -2347 |
| 90 | -586 | -493 | -1079 |
| 95 | -363 | 264 | -99 |
| 110 | 331 | 2162 | 2493 |
| 125 | 789 | 3416 | 4205 |

Table 4. *d-d* and *s-d* contributions (in cm$^{-1}$) to $H_{AB}$ as a function of the bonding distance between Cu and the bridging ligands calculated with mon1.

| $d_{Cu\text{-}F}$ [Å] | *d-d* | *s-d* |
|---|---|---|
| 1.7 | 1939 | -7318 |
| 1.8 | 1109 | -3913 |
| 1.88 | 710 | -2361 |
| 1.94 | 503 | -1632 |
| 2 | 357 | -1097 |
| 2.1 | 210 | -575 |
| 2.2 | 120 | -298 |

Table 5. Quantities required for calculating $H_{AB}$ with eq.(9) for three different bridging angles with the methods mon2, mon3.

|  | 80° | | 90° | | 120° | |
|---|---|---|---|---|---|---|
|  | mon2 | mon3 | mon2 | mon3 | mon2 | mon3 |
| $\varepsilon_A^{mono}$ (eV) | -10.371 | -9.850 | -10.306 | -9.791 | -10.260 | -9.732 |
| $H_{AB}^{mono}$ (eV) | -0.094 | -0.014 | 0.024 | 0.023 | -0.030 | -0.002 |
| $S_{AB}^{mono}$ | 0.002 | 0.007 | 0.002 | 0.001 | -0.006 | -0.008 |
| $H_{AB}$ (cm$^{-1}$) | 555 | 424 | 321 | 298 | -733 | -653 |

Table 6. $H_{AB}$, $J(bs)$, and $J(fit)$ from FPLO where $C = 110$ and $f = 1/7000$ (in cm$^{-1}$).

| $\theta$ [°] | $H_{AB}$ | $J(bs)$ | $J(fit)$ |
|---|---|---|---|
| 82.5 | -46 | 110 | 109 |
| 90 | 475 | 67 | 46 |
| 95 | 889 | -82 | -116 |
| 110 | 1985 | -1016 | -1016 |
| 125 | 2203 | -1292 | -1277 |

**FIGURES**

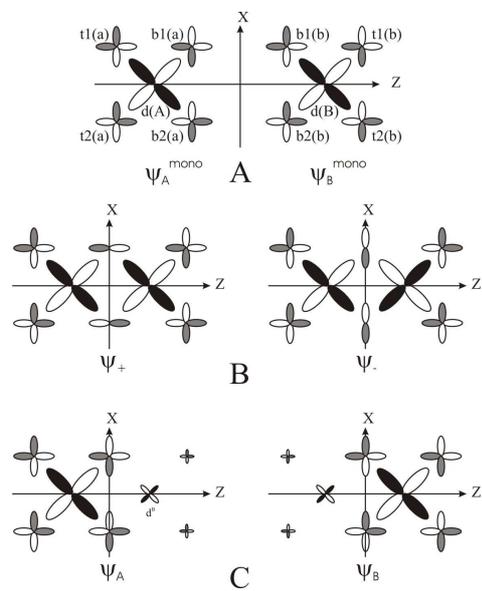

Figure 1.

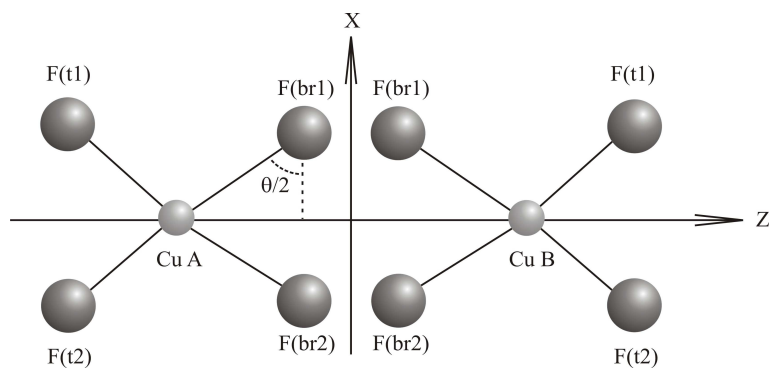

Figure 2.

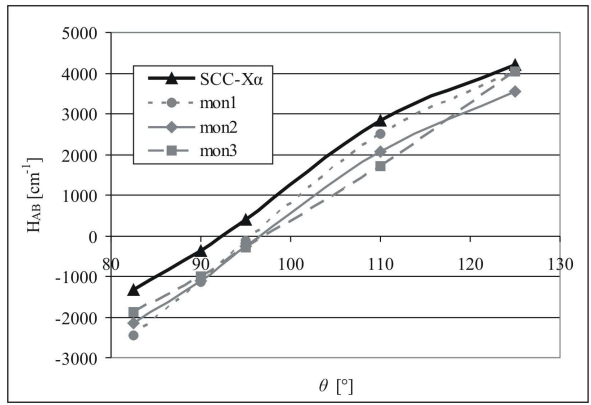

Figure 3.

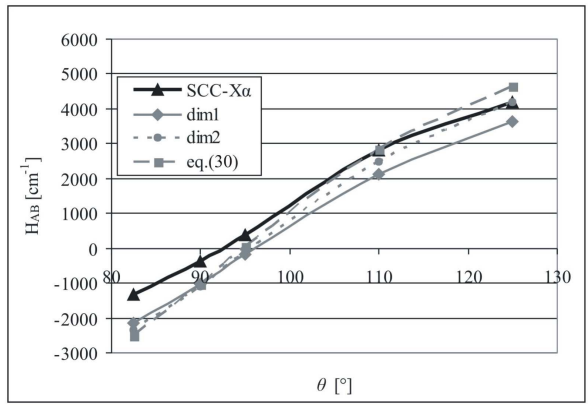

Figure 4.

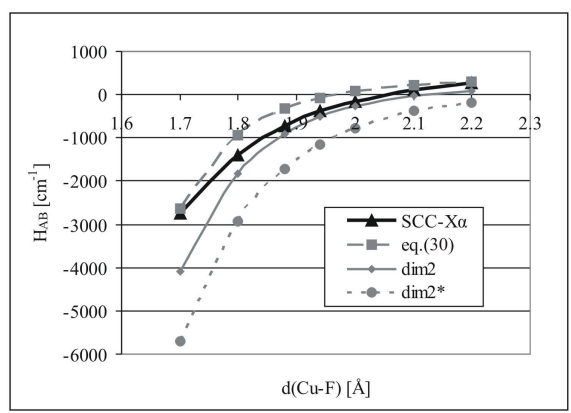

Figure 5.

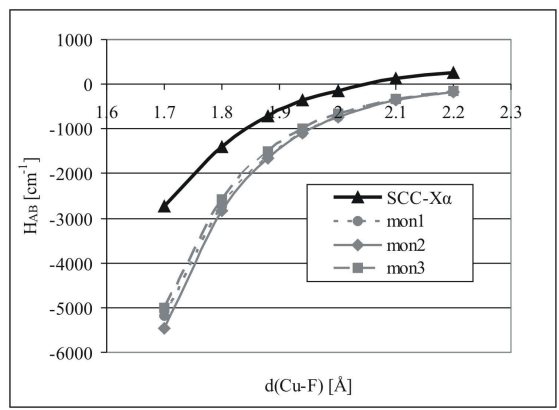

Figure 6.

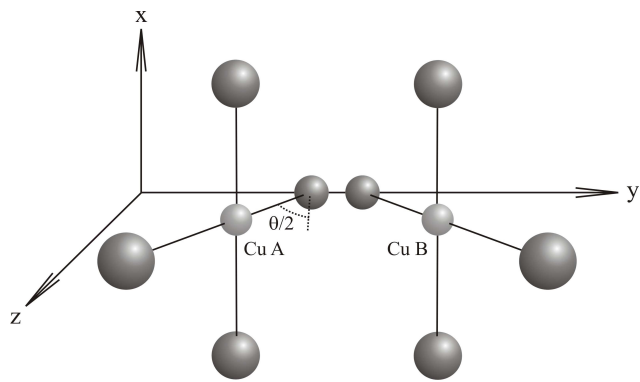

Figure 7.

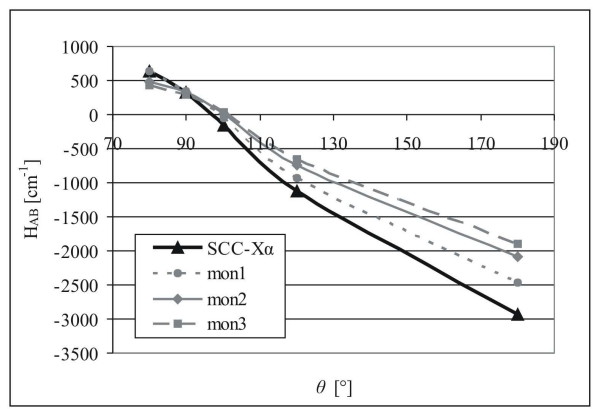

Figure 8.

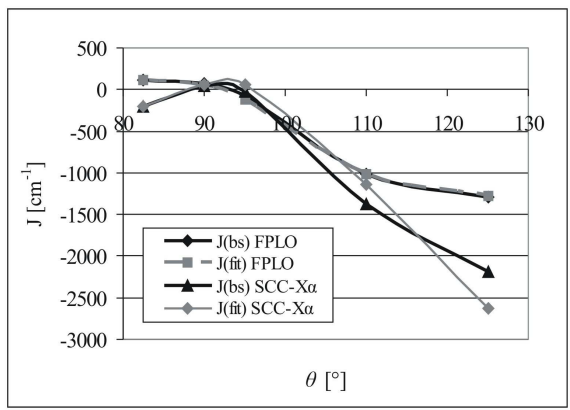

Figure 9.

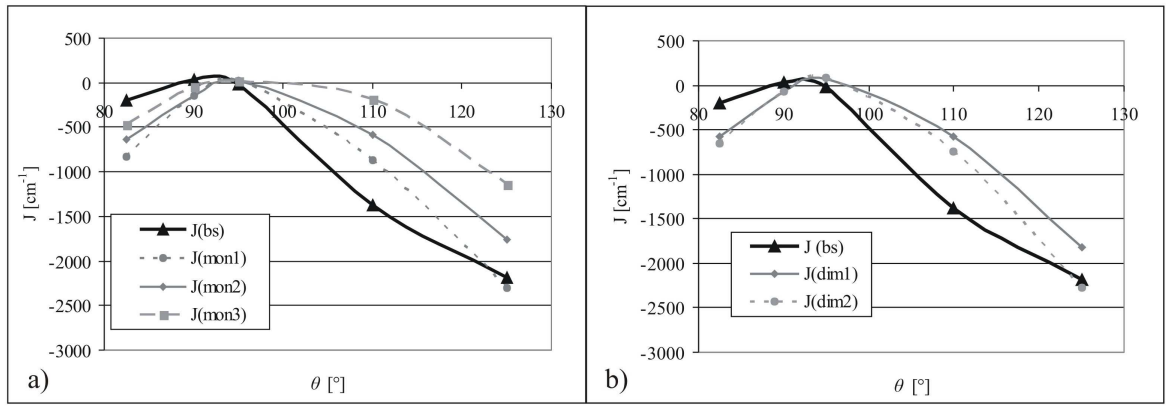

Figure 10.

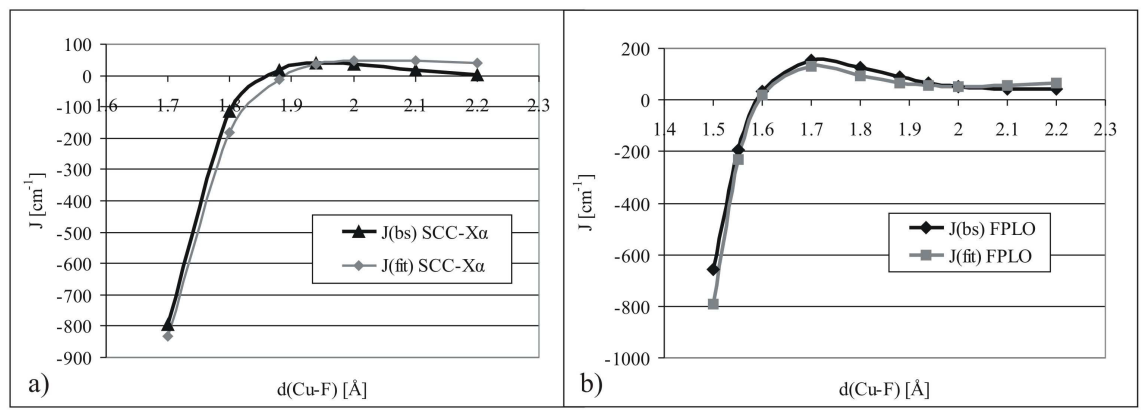

Figure 11.

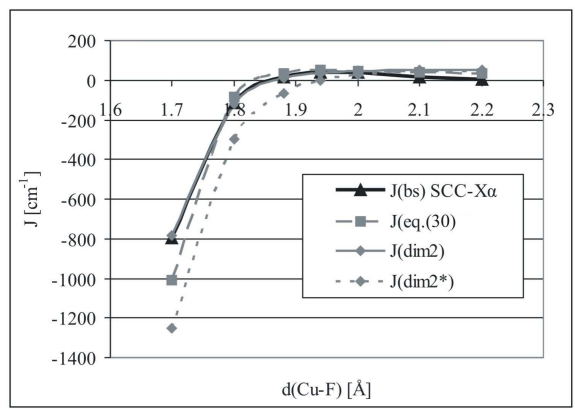

Figure 12.

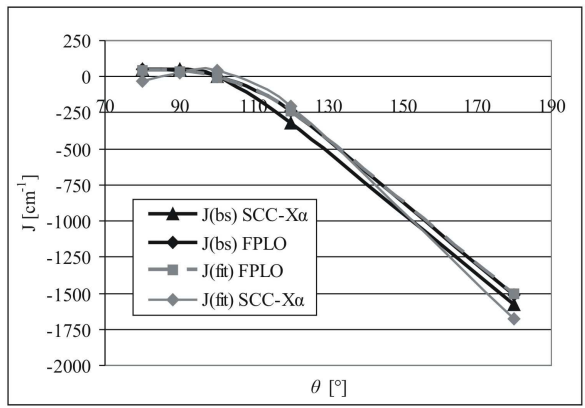

Figure 13.

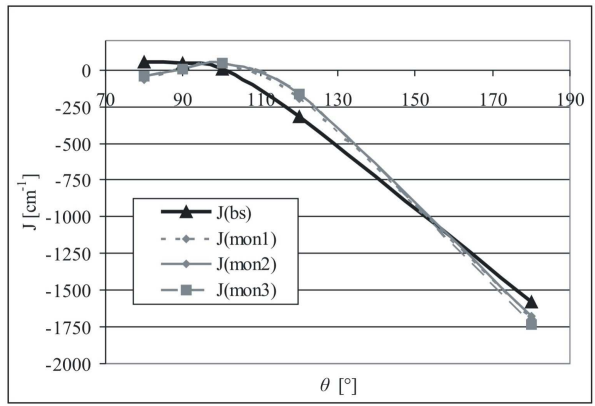

Figure 14.

**FIGURE CAPTIONS**

Figure 1. Construction of localized MO's of a planar, symmetrical, doubly bridged dimer placed in the $xz$-plane. $d(X)$ is a transition metal d-orbital on monomer centre $X$. t1($x$), t2($x$), b1($x$) and b2($x$) are the $p_z$- and $p_x$-orbitals of the terminal (t) and bridging (b) ligands, respectively, on centre $x$ (for ligand orbitals the centre is written in lower case). The shadings indicate the phases of the wavefunctions and exhibit the antibonding character of the MO's $\psi_X^{mono}$. The ligand $s$-orbitals are omitted. The size of the orbitals is proportional to their contribution to the respective MO. A: Decomposition of the dimer into two monomers and construction of the two singly occupied monomer MO's. In contrast to the $p_z$-orbitals, the $p_x$-orbitals of the bridging ligands of the two monomers have different phases. B: The delocalized HOMO and LUMO of the dimer; C: The localized singly occupied MO's;

Figure 2. Decomposition of the $\left[ Cu_2 F_6 \right]^{2-}$ complex into two monomers. $t$ and $br$ are terminal and bridging ligands, respectively.

Figure 3. Comparison of the numerical $H_{AB}$ (in cm$^{-1}$) from SCC-X$\alpha$ with the three monomer approaches (mon1-mon3) for different bridging angles $\theta$.

Figure 4. Comparison of the numerical $H_{AB}$ (in cm$^{-1}$) from SCC-X$\alpha$ with the two dimer approaches (dim1 and dim2) for different bridging angles $\theta$.

Figure 5. $H_{AB}$ (in cm$^{-1}$) as a function of the bonding distance between Cu and the bridging ligands calculated with the dimer methods, dim2 and eq.(30), and compared with the numerical values from SCC-X$\alpha$ calculations. dim2* is the value without ligand-ligand interactions.

Figure 6. $H_{AB}$ (in cm$^{-1}$) as a function of the bonding distance between Cu and the bridging ligands calculated with different monomer-methods, without ligand-ligand interactions, compared with the numerical values from SCC-X$\alpha$.

Figure 7. Singly bridged $\left[Cu_2F_7\right]^{3-}$ complex, decomposed into two monomers of symmetry $C_{2v}$.

Figure 8. Comparison of $H_{AB}$ (in cm$^{-1}$) from SCC-X$\alpha$ and the three monomer approaches, respectively, for different bridging angles.

Figure 9. Comparison between $J(bs)$ and the fitted $J(fit)$ from SCC-X$\alpha$ ($C = 80$ and $f = 1/13000$) and FPLO ($C = 80$ and $f = 1/8500$) calculations. (in cm$^{-1}$)

Figure 10. Comparison between $J(bs)$ from SCC-X$\alpha$ and $J(analyt)$, in cm$^{-1}$, for $H_{AB}$ taken from the (a) monomer ($f = 1/7000$ and $C = 30$) and (b) dimer approach ($f = 1/15000$ and $C = 80$).

Figure 11. Comparison between $J(bs)$ and the fitted $J(fit)$ from (a) SCC-X$\alpha$, $C = 50$, $f = 1/17000$, and (b) FPLO, $C = 130$, $f = 1/6000$ (in cm$^{-1}$).

Figure 12. Comparison between $J(bs)$ and $J(analyt)$ for $H_{AB}$ from the dimer approach dim2 ($f = 1/40000$, $C = 50$), the same approach without ligand-ligand interactions dim2* ($f = 1/50000$, $C = 50$) and the diagonalization with perturbation calculation eq.(30) ($f = 1/13000$, $C = 50$).

Figure 13. $J(bs)$ and the fitted $J(fit)$ from SCC-X$\alpha$ ($C = 50$, $f = 1/10000$) and from FPLO ($C = 80$ and $f = 1/8500$) calculations.

Figure 14. Comparison between $J(bs)$ from SCC-X$\alpha$ and $J(analyt)$ for $H_{AB}$ from the monomer approaches (mon1: $f = 1/6000$, $C = 50$;  mon2: $f = 1/5000$, $C = 50$; mon3: $f = 1/4000$, $C = 50$).

**GRAPHICAL ABSTACT**

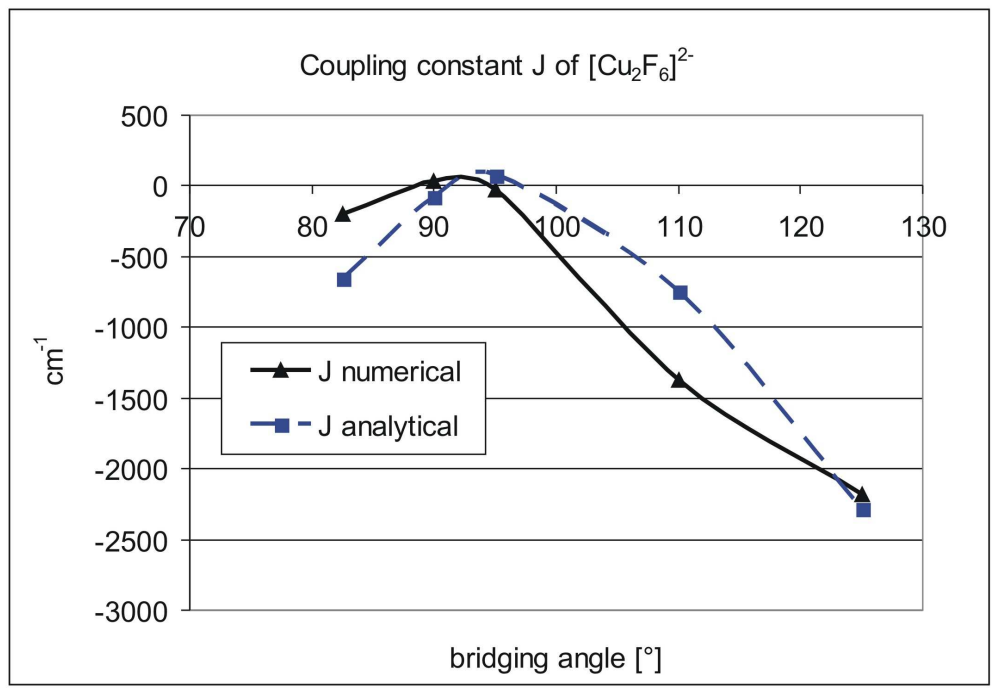

Coupling constant J of $[Cu_2F_8]^{2-}$